\documentclass[a4paper,12pt]{article}
\usepackage[utf8]{inputenc}

\usepackage{amsmath,amsfonts,amsthm,amssymb,dsfont}
\usepackage{graphicx,wrapfig,lipsum}
\usepackage{layout}
\usepackage[section]{placeins}
\usepackage{indentfirst}

\usepackage{tikz}
\usetikzlibrary{shapes.geometric,decorations.markings}
\usepackage{subfigure}

\usepackage[hidelinks]{hyperref}
\hypersetup{colorlinks = true, linkcolor = red, linktocpage = true, citecolor = blue}

\usepackage{geometry}
\numberwithin{equation}{section}
\newcommand{\N}{\mathcal{N}}
\newcommand{\Nc}{N_{c}}
\newcommand{\Nf}{N_{f}}
\newcommand{\tNc}{\tilde{\Nc}}
\newcommand{\tPhi}{\tilde{\Phi}}
\newcommand{\tpsi}{\tilde{\psi}}
\newcommand{\tlambda}{\tilde{\lambda}}

\def\beq{\begin{equation}}
\def\eeq{\end{equation}}
\newcommand{\bea}{\begin{eqnarray}}
\newcommand{\eea}{\end{eqnarray}}
\def\bal{\begin{align}}
\def\eal{\end{align}}

\begin{document}

\begin{titlepage}

\begin{center}

$\phantom{.}$\\ \vspace{2cm}
\noindent{\Large{\textbf{Dualities of 3D ${\cal N}=1$ SQCD from Branes and non-SUSY deformations}}}

\vspace{1cm}

Adi Armoni \footnote{a.armoni@swansea.ac.uk} and Ricardo Stuardo \footnote{ricardostuardotroncoso@gmail.com}

\vspace{0.5cm}

\textit{Department of Physics, Faculty of Science and Engineering\\
        Swansea University, SA2 8PP, UK}

\end{center}

\vspace{0.5cm}
\centerline{\textbf{Abstract}} 

\vspace{0.5cm}

\noindent{We study the dynamics of an 'electric' ${\cal N}=1$ 3D $U(N_c)_{k,k+\frac{N_c}{2}}$ SQCD theory. By embedding the theory in string theory, we propose that the theory admits a 'magnetic' dual and analyse the low energy dynamics of the theory using its dual. When $\frac{N_f}{2} \ge\frac{N_c}{2}-k$ the IR dynamics is described by either a TQFT for large quark masses, or a Grassmannian and a Wess-Zumino (WZ) term for small masses. We also consider non-supersymmetric mass deformations and RG flows in the vicinity of the SUSY point and find agreement between the IR of the electric and its magnetic dual. When $\frac{N_f}{2} < \frac{N_c}{2}-k$ supersymmetry is broken and the IR dynamics is a described by a TQFT accompanied by a Goldstino. We also discuss SQCD theories based on $SO$/$USp$ gauge groups. }

\vspace*{\fill}

\end{titlepage}

\newpage

\tableofcontents

\thispagestyle{empty}

\newpage
\setcounter{page}{1}
\setcounter{footnote}{0}

\section{Introduction}

Three dimensional gauge theories with a Chern-Simons (CS) term, with or without supersymmetry, attracted a lot of attention in recent years, following \cite{Aharony:2008ug,Aharony:2012nh}. Thanks to the CS term those theories admit a rich structure where the IR dynamics exhibits phases with a CFT, or a gapped phase, a broken flavour symmetry phase, etc. Often the IR theory contains a TQFT. The IR dynamics is constrained by matching global anomalies.

Duality plays a central role in the study of strongly coupled 3D gauge theories. It is by now clear that most of these dualities can be thought of as Seiberg dualities and can be better understood by embedding them in string theory \cite{Giveon:2008zn}, even in the absence of supersymmetry \cite{Armoni:2017jkl}.

In this paper we study gauge theories which contain one adjoint fermion and $N_f$ flavours of quarks. The theories we consider are embedded in ${\cal N}=1$ SQCD and can be realised by Hanany-Witten brane configurations, similar to that used by Giveon and Kutasov \cite{Giveon:2008zn}. By using the brane picture we propose dualities that can be checked by explicit field theory computations. In particular, we can verify that the Witten index of the electric theory and its magnetic dual is the same and that both theories flow in the IR to the same fixed point. Another advantage of the string theory embedding is that the matching of global anomalies is guaranteed.

The prime gauge theory we consider is 3D 'electric' $U(N_c)_{k,k+\frac{N_c}{2}}$ SQCD with $N_f$ flavours, which we call quarks (the content of each SUSY multiplet is described in section \ref{sec2}). We propose that for $\tilde N_c \equiv \frac{N_f}{2}+k -\frac{N_c}{2} \ge0$ the 'electric' theory admits a $U(\frac{N_f}{2}+k -\frac{N_c}{2})_{-\frac{k}{2}-\frac{3N_c}{4} + \frac{N_f}{4},-k-\frac{N_c}{2}}$ SQCD 'magnetic' dual with $N_f$ quarks and a WZ term. The IR theory contains two phases: the large quark mass, $|m_{\psi}|\gg g^{2}k$ phase, where the IR theory is a TQFT and the small quark mass phase, $|m_{\psi}|\ll g^{2}k$ . The latter phase consists of sectors labelled by $n$ where the IR theory is ${\cal N}=1$ $U(\frac{N_f}{2}+k -\frac{N_c}{2} -n)$ SQCD accompanied by a Grassmannian and a WZ term.
The pattern of flavour symmetry breaking is 
 \beq
 U(N_f) \rightarrow  U(N_f -n) \times U(n) \, .
 \eeq
 
 The WZ term is 
\beq
 N_c \int _{\cal M}  {\rm tr}\, \frac{1}{2\pi} F \wedge F \, ,
 \eeq
where ${\cal \partial M}$ is the 3D spacetime. $F$ transforms in the adjoint of $U(N_f-n)$. It is important to note that in the electric theory there is no WZ term in the UV, which, as we will discuss in section \ref{sec2}, can be seen directly from the fact that in the brane set-up the flavour branes end on the $NS5$ brane instead of the $(1,k')$ fivebrane, which is the case for the magnetic theory. This is similar to Seiberg Duality, where the meson operator can only be seen in the magnetic theory. We expect the WZ term to appear in the IR of the electric theory, since its the same as the IR of the magnetic counterpart.

When  $\tNc \equiv \frac{N_f}{2}+k -\frac{N_c}{2} < 0$ SUSY is broken and the 'electric' theory is described by two magnetic theories: a magnetic dual of the form $U(\frac{N_c}{2}-\frac{N_f}{2}-k)_{\frac{k}{2}+\frac{3N_c}{4} -\frac{N_f}{4},k+\frac{N_c}{2}}$ and magnetic' of the form $U(\frac{N_c}{2}-\frac{N_f}{2}+k)_{\frac{k}{2}-\frac{3N_c}{4} +\frac{N_f}{4},k-\frac{N_c}{2}}$. The IR is described by several TQFTs.

This work is a continuation of  \cite{Bashmakov:2018wts,Gomis:2017ixy,Armoni:2022wef}, where the case of adjoint QCD without flavours was considered. Here we give a string theory perspective on the phenomena that occurs due to the addition of flavours, in particular the symmetry breaking phase and the appearance of a WZ term. When the adjoint fermion is integrated out, we arrive at QCD. Our results agree with \cite{Komargodski:2017keh}.
Earlier field theory papers considered the anomalies of the theory when $k=0$ \cite{Lohitsiri:2022jyz} and the SQCD duality \cite{Benini:2018umh,Choi:2018ohn,Bashmakov:2018ghn}. We also provide evidence for new IR dualities of theories where SUSY is spontaneously broken.

By adding $O3$ planes to the brane configuration we realise $SO$ and $Sp$ gauge theories and obtain the associated dualities. 

\section{\texorpdfstring{$\N=1$}{N=1} Dualities from Branes}\label{sec2}

We derive the gauge theory dualities from a duality in string theory. The electric (or magnetic) configuration consist of $\Nc$ (or $\tNc$) $D3$-branes suspended between an $NS5$-brane and a $(1,k')$ tilted fivebrane. The D3 branes span the 012 directions and an interval in the 6 direction. The $NS5$-brane spans the 012345 directions. The $(1, k')$ fivebrane spans the 01238 directions and it is tilted in the (59) plane. In order to obtain $U(\Nf)$ flavour symmetry, we add $\Nf$ semi-infinite $D3$-branes ending on the $NS5$-brane on the electric theory (or the $(1,k')$  fivebrane in the magnetic one). This set-up is the one considered in \cite{Giveon:2008zn} and realises 3D $\N=2$ $U(\Nc)_{k',k'}$ Yang-Mills-Chern-Simons (YM-CS) theory. The electric and magnetic configurations are as in the figure below
    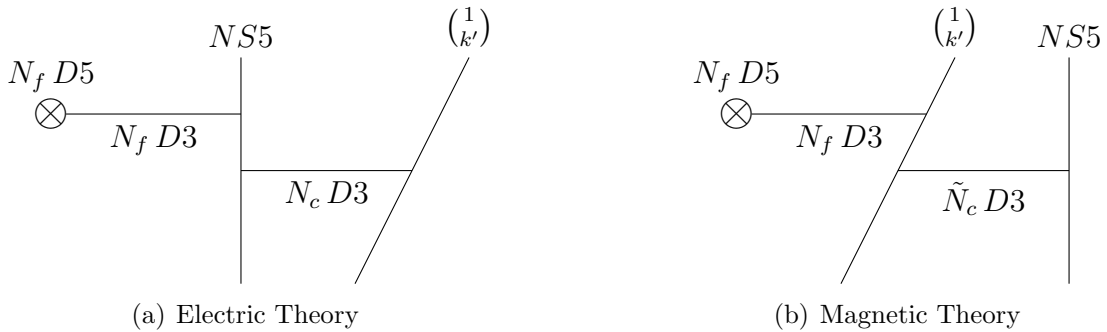
\begin{figure}[h!]
    \centering
    
    %
    \subfigure[Electric Theory]{
    \begin{tikzpicture}
    
    \draw (-1.5,-1.5) -- (-1.5,1.5) node [above] {$NS5$};

    \draw (-1.5,0) -- (0.75,0) node [midway,below] {$N_{c} \, D3$};

    \draw (0,-1.5) -- (1.5,1.5) node [above] {$\binom{1}{k'}$};

    \draw (-3.8,0.75) -- (-1.5,0.75) node [midway,below] {$N_{f} \, D3$}; 
    
    \draw (-4,0.9) node[above] {$\Nf \, D5$};
    \node at (-4,0.75){\Large $\otimes$};
   
    \end{tikzpicture}}
    \quad\quad\quad\quad\quad%
    %
    %
    \subfigure[Magnetic Theory]{
    \begin{tikzpicture}
    \draw (1.5,-1.5) -- (1.5,1.5) node [above] {$NS5$};

    \draw (-0.75,0) -- (1.5,0) node [midway,below] {$\tilde{N}_{c} \, D3$};

    \draw (-1.5,-1.5) -- (0,1.5) node [above] {$\binom{1}{k'}$};

    \draw (-2.675,0.75) -- (-0.375,0.75) node [midway,below] {$N_{f} \, D3$};
    
    \draw (-2.875,0.9) node[above] {$\Nf \, D5$};
    \node at (-2.875,0.75){\Large $\otimes$};
    
	\end{tikzpicture}
	\label{MagSetUp}
    }
    \caption{Dualities from Branes}
    \label{BraneConfiguration}
	\end{figure}

Both $\N=2$ electric and magnetic theories can be written in terms of $\N=1$ variables as follows: The $\N=2$ vector multiplet splits into a $\N=1$ vector multiplet $(A_{\mu},\lambda)$ and an adjoint scalar multiplet $(\chi,\varphi)$, while the $\N=2$ flavours split into two $\N=1$ chiral multiplets $\Phi$ and $\Phi'$ (here $\Phi = (\psi,\phi)$, similarly for $\Phi'$).

Let us begin with the electric theory. In order to obtain 3D $\N=1$ $U(\Nc)$ with one adjoint fermion and $\Nf$ flavours as the low energy theory, we set\footnote{We could also start with level $k'' = k - \frac{\Nc}{2} + \frac{\Nf}{2}$, with $k''<0$, and give masses to the fields as $m_{\chi}>0$ and $m_{\Phi'}<0$. This defines the electric' theory and its dual magnetic'. When the theories are SUSY, the range of $k$ is different: $k>\frac{\Nc-\Nf}{2}$ for the electric-magnetic and $k<-\frac{\Nc-\Nf}{2}$ for the electric'-magnetic'. The pairs of theories are related by parity. In the non-SUSY case, $|k|< \frac{\Nc-\Nf}{2}$ in both cases. We conjecture that in the non-SUSY case, the electric and electric' theories are the same since the $U(1)$ factor decouples. Then, the magnetic and magnetic' allows us to describe three different phases of the $SU(\Nc)_{k}$ electric theory. } $k' = k + \frac{\Nc}{2} - \frac{\Nf}{2}$ (with $(k'>0)$) and then integrate out $\chi$ for large negative mass, which shifts the $SU(\Nc)$ CS level by $-\frac{\Nc}{2}$,  and $\tpsi'$, shifting both CS-levels by $\frac{\Nf}{2}$. This can be achieved by further rotation of the tilted fivebrane as in \cite{Armoni:2022wef}. We then obtain 3D $\N=1$ $U(\Nc)_{k,k + \frac{\Nc}{2}}$ with $\Nf$ flavours $\Phi$. 

To obtain the Seiberg dual, we swap the fivebranes following the prescription of \cite{Giveon:2008zn,Elitzur:1997fh}. In order to preserve supersymmetry we consider the case $\tilde N_c \ge 0$. After swapping the fivebranes we obtain the $\N=1$ magnetic theory  $U(\tNc)_{\tilde{k},-k'+\frac{\Nf}{2}}$, where the number of $D3$ branes on the magnetic side is
    \begin{equation}
        \tNc = k'-\Nc+\Nf =  k - \frac{\Nc}{2} + \frac{\Nf}{2}.
    \end{equation}

As in the electric side, the $SU(\tNc)$ CS level is shifted by the massive adjoint fermion and the massive chiral flavour\footnote{In the magnetic theory, the masses of the adjoint fermion and the flavours have opposite signs when compared with their electric counterparts.}
    \begin{equation}
        \tilde{k} = -k' + \frac{\tNc}{2} - \frac{\Nf}{2} = -\frac{k}{2} - \frac{3\Nc}{4} + \frac{\Nf}{4},
    \end{equation}
while the $U(1)$ CS-level only receives a contribution from the flavours $-k'+\frac{\Nf}{2} = -k -\frac{\Nc}{2}$. Therefor we propose the following Seiberg duality between two $\N=1$ supersymmetric YM-CS theories
    \begin{equation}
      U(\Nc)_{k,k + \frac{\Nc}{2}} + \Nf \, \Phi
      \Leftrightarrow  
      U\left(  k - \frac{\Nc}{2} + \frac{\Nf}{2} \right)_{-\left(\frac{k}{2} + \frac{3\Nc}{4} - \frac{\Nf}{4}\right),-\left(k+\frac{\Nc}{2}\right)} + \Nf \, \tPhi,
      \label{N1magnetic}
    \end{equation}
where we have denoted the flavours in the magnetic theory by $\tPhi$. This duality was obtained from the field theory side in \cite{Benini:2018umh,Choi:2018ohn}. Note that the r.h.s. of \eqref{N1magnetic}, namely the magnetic theory, admits also mesons and an additional CS term, arising from the intersection of the flavour branes with the tilted fivebrane which is interpreted as a WZ term in the phase where the flavour symmetry is broken.

\subsection{Witten Index}

To check the duality, we derive the Witten Index from the brane configuration. We need to count all different SUSY vacua as in \cite{Giveon:2008zn}. On the electric side, it is convenient to move the $D5$-brane past the $NS5$-brane and set them on top of the $D3$-branes. On a SUSY vacuum, we allow $n$ of the $\Nc$ $D3$-branes to break on the $D5$-brane, so that half of the $n$ $D3$ stretches between the $NS5$ and the $D5$, and the other half stretches between the $D5$ and the $(1,k')$ fivebrane. Sending the $n$ $D5$s to infinity leads to a brane configuration with $\Nc\rightarrow \Nc-n$ and $\Nf\rightarrow \Nf-n$. Then we need to count how the leftover $\Nc-n$ $D3$-branes can end on the $(1,k')$ brane. Summing over all possible $n$ leads to
    \begin{equation}
      I_{W} = \sum^{\Nc}_{n=0} \binom{k'}{\Nc-n} \binom{\Nf}{n} = \binom{k'+\Nf}{\Nc} 
        = \frac{\left(k+\frac{\Nc}{2}+\frac{\Nf}{2}\right)!}{\Nc!\left(k-\frac{\Nc}{2}+\frac{\Nf}{2}\right)!}.
    \end{equation}

On the magnetic side the argument is similar. We allow $n$ of the $\tNc$ $D3$-branes to reconnect with $n$ of the $\Nf$ $D3$ flavour branes. Then we count how the leftover $\tNc-n$ branes can end on the tilted fivebrane
    \begin{equation}
        I_{W} = \sum^{\tilde \Nc}_{n=0} \binom{k'}{\tNc-n} \binom{\Nf}{n} = \binom{k'+\Nf}{\tNc} 
        = \frac{\left(k+\frac{\Nc}{2}+\frac{\Nf}{2}\right)!}{\Nc!\left(k-\frac{\Nc}{2}+\frac{\Nf}{2}\right)!}.
    \end{equation}
 
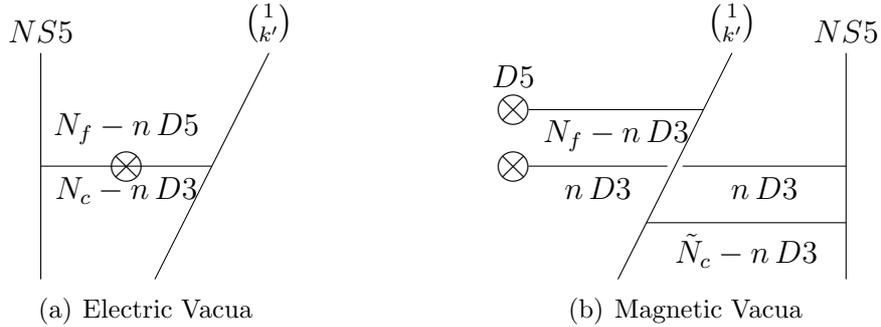
\begin{figure}
    \centering
    
    %
    \subfigure[Electric Vacua]{
    \begin{tikzpicture}
    
    \draw (-1.5,-1.5) -- (-1.5,1.5) node [above] {$NS5$};

    \draw (-1.5,0) -- (0.75,0) node [midway,below] {$N_{c}-n \, D3$};

    \draw (0,-1.5) -- (1.5,1.5) node [above] {$\binom{1}{k'}$};
    
    \draw (-0.375,0.15) node[above] {$\Nf-n \, D5$};
    \node at (-0.375,0){\Large $\otimes$};
   
    \end{tikzpicture}}
    \quad\quad\quad\quad\quad%
    %
    %
    \subfigure[Magnetic Vacua]{
    \begin{tikzpicture}
    \draw (1.5,-1.5) -- (1.5,1.5) node [above] {$NS5$};

    \draw (-0.65,0) -- (1.5,0) node [midway,below] {$n \, D3$};
    \draw (-1.125,-0.75) -- (1.5,-0.75) node [midway,below] {$\tilde{N}_{c}-n \, D3$};

    \draw (-1.5,-1.5) -- (0,1.5) node [above] {$\binom{1}{k'}$};

    \draw (-2.675,0.75) -- (-0.375,0.75) node [midway,below] {$N_{f}-n \, D3$};
    \draw (-2.675,0) -- (-0.85,0) node [midway,below] {$n \, D3$};
    
    \draw (-2.875,0.9) node[above] {$D5$};
    \node at (-2.875,0.75){\Large $\otimes$};
    \node at (-2.875,0){\Large $\otimes$};

	\end{tikzpicture}
	\label{SUSYvacuumMag}
    }
    \caption{SUSY vacua on both sides of the duality}
    \label{SUSYvacuum}
	\end{figure}

Both $n$th SUSY vacua are shown in figure \ref{SUSYvacuum}.
	
\subsection{Low energy with  Symmetry Breaking}
\label{SSB}

When  $\tilde N_c \equiv \frac{N_f}{2}+k -\frac{N_c}{2} \ge0$ the dual SQCD theory admits two phases: large and small quark mass. In this section we discuss the small mass phase.

Let us consider the brane configuration of the magnetic theory, depicted in Figure \ref{MagSetUp}. When the quark mass is small, the flavour branes are located near the colour branes along the tilted fivebrane axis. Therefore colour and flavour can be reconnected and form vacua with three kind of D3 branes: leftover colour branes, leftover flavour branes and reconnected branes. In a vacuum where $n$ branes are reconnected we have $N_f -n$ leftover flavour branes and $\tilde N_c - n$ leftover colour branes, see Figure \ref{SUSYvacuumMag}. In field theory those vacua correspond to a dual squark condensation
\beq
\langle \tilde \phi ^i _\alpha  \rangle \ne 0 \,
\eeq
where $i$ and $\alpha$ are flavour and dual colour indices, respectively.

The supersymmetric field theory that lives on the brane configuration consists of the following ingredients: $U(\tilde N_c -n)$ SQCD theory with a broken flavour symmetry of the following pattern
\beq
U(N_f) \rightarrow U(N_f-n) \times U(n)
\eeq
described by the Grassmannian $U(N_f) / U(N_f-n) \times U(n)$. The $2n(N_f-n)$ Nambu-Goldstone bosons correspond to the open strings stretched between the stack of $n$ $D3$-branes and the stack of $N_f-n$ $D3$-branes. These vacua were described in \cite{Benini:2018umh,Choi:2018ohn}.

In addition to the Grassmannian there is a WZ term, as argued in \cite{Komargodski:2017keh}. We claim that the source of this term in the brane setup is a CS term that lives on the intersection of the leftover flavour branes and the tilted fivebrane. Indeed, as was recently argued in \cite{Armoni:2022xhy} when a $D3$-brane is suspended between a $D5$-brane and a tilted fivebrane there is a CS term living on the intersection, even when the $D5$-brane is taken to infinity \cite{Gaiotto:2008ak}.

The WZ term is written as
\beq
 k_f \int _{\cal M}  {\rm tr}\, \frac{1}{2\pi} F \wedge F \, ,
 \eeq
 where $F$ transforms in the adjoint of $U(N_f-n)$. Let us calculate the level $k_f$. It is the level $k'$ shifted by the fermions on the flavours branes and colour branes
 \beq
 k_f = k' + (N_f -n) - (\tilde N_c -n) = k' + N_f - \tilde N_c = N_c
 \eeq
 namely, in each vacuum of the magnetic theory there is a WZ term of the form
\beq
 N_c \int _{\cal M}  {\rm tr}\, \frac{1}{2\pi} F \wedge F \, .
 \eeq
The gauge field $F=dA$ is not independent of the mesons, but as in the standard chiral Lagrangian, $A\sim M^\dagger \partial  M$.

 When supersymmetry is broken by giving a small mass to the adjoint fermion, the vacuum degeneracy is lifted and one vacuum is selected. It is natural to propose that the colour group disappears, namely the chosen vacuum is $n=\tilde N_c$. The motivation is Coleman-Witten theorem in 3D: flavour symmetry may be broken to $U(m) \times U(N_f-m)$ without a Yang-Mills gauge theory \cite{Armoni:2019lgb}. Hence, in the vacuum with  $n=\tilde N_c$ we have symmetry breaking of the form
 \beq
    U(N_f) \rightarrow U\left(\frac{N_f}{2}- k + \frac{N_c}{2} \right)  \times 
            U \left(\frac{N_f}{2} + k - \frac{N_c}{2} \right) \, .
 \eeq
 
 The WZ term is 
\beq
 N_c \int _{\cal M}  {\rm tr}\, \frac{1}{2\pi} F \wedge F \, ,
 \eeq
 where ${\cal \partial M}$ is the 3D spacetime. $F$ transforms in the adjoint of $U(\frac{N_f}{2} -k +\frac{N_c}{2})$.

 Note that if we integrate out the adjoint fermion in the electric side $k \rightarrow k-\frac{N_c}{2}$ and our proposal for a symmetry breaking pattern and the Grassmannian coincide with the proposal of \cite{Komargodski:2017keh} for QCD.
 
While we consider the case $U(\Nc)_{k,k+\frac{\Nc}{2}}$, note that the $SU(\Nc)_{k}$ is special when $k=0$, since in this case the latter theory preserves parity, so that a flavour symmetry breaking of the form $U(N_f) \rightarrow U(\frac{N_f}{2}+ \frac{N_c}{2})  \times U(\frac{N_f}{2} - \frac{N_c}{2})$ is forbidden, due to Vafa-Witten theorem. We propose that in the $U(\Nc)$ case with $k=0$ squark condensation does not occur and the theory is described by a TQFT, as we discuss in the next section.

\section{Low Energy Dynamics with a TQFT}

Let us consider the case of large quark mass ($m_{\psi}$). In the case with $\Nf=0$, the electric theory is supersymmetric when the adjoint fermion mass is $m_{\lambda}=-g^{2}k$. In the deep IR, this theory admits two phases described by two TQFTs. It was shown that using the magnetic dual, one is able to recover the TQFT in a neighbourhood of a SUSY point. Here we show that this match of the IR dynamics around the SUSY point also holds in the presence of flavours.

In order to obtain the IR dynamics of the electric theory, we need to integrate out the adjoint fermion $\lambda$ and the flavours $\psi$. There are four possible TQFTs in the IR depending on the sign of the masses $m_{\lambda}$ and $m_{\psi}$. These theories are listed in the table below
    \begin{table}[h!]
        \centering
        \begin{tabular}{|c|c|c|}\hline
                       &        $m_{\lambda}<0$            &        $m_{\lambda}>0$         \\ \hline
             $m_{\psi}>0$ & $U(\Nc)_{k - \frac{\Nc}{2} + \frac{\Nf}{2} , k + \frac{\Nc}{2} + \frac{\Nf}{2}}$  &   $U(\Nc)_{k + \frac{\Nc}{2} + \frac{\Nf}{2} , k + \frac{\Nc}{2} + \frac{\Nf}{2}}$ \\ \hline
             $m_{\psi}<0$ & $U(\Nc)_{k - \frac{\Nc}{2} - \frac{\Nf}{2} , k + \frac{\Nc}{2} - \frac{\Nf}{2}}$  &   $U(\Nc)_{k + \frac{\Nc}{2} - \frac{\Nf}{2} , k + \frac{\Nc}{2} - \frac{\Nf}{2}}$ \\ \hline
          \end{tabular}
          \caption{Possible IR theories on the electric theory}
          \label{elec}
      \end{table}

Repeating the same analysis on the magnetic side leads to the following TQFTs
    \begin{table}[h!]
        \centering
        \begin{tabular}{|c|c|c|}\hline
                       &          $m_{\tlambda}<0$       &   $m_{\tlambda}>0$   \\ \hline
             $m_{\tpsi}>0$ &  $U\left(  k - \frac{\Nc}{2} + \frac{\Nf}{2} \right)_{ -k - \frac{\Nc}{2} + \frac{\Nf}{2}, -k - \frac{\Nc}{2} + \frac{\Nf}{2}}$  
                       & $U\left(  k - \frac{\Nc}{2} + \frac{\Nf}{2} \right)_{ -\Nc+\Nf, -k - \frac{\Nc}{2} + \frac{\Nf}{2}}$ \\ \hline
             $m_{\tpsi}<0$ &  $U\left(  k - \frac{\Nc}{2} + \frac{\Nf}{2} \right)_{ -k - \frac{\Nc}{2} - \frac{\Nf}{2}, -k - \frac{\Nc}{2} - \frac{\Nf}{2}}$  
                       & $U\left(  k - \frac{\Nc}{2} + \frac{\Nf}{2} \right)_{-\Nc,-k - \frac{\Nc}{2} - \frac{\Nf}{2}}$ \\ \hline
        \end{tabular}
        \caption{Possible IR theories on the magnetic theory}
        \label{mag}
    \end{table}

Since from level-rank duality\footnote{Level-Rank Duality for $U-U$ theories is given by \cite{Hsin:2016blu}
    \begin{equation*}
        U(N)_{K,K\pm N} \Longleftrightarrow U(K)_{-N,-N\mp K}
    \end{equation*}}
we have
    \begin{equation}
       U(\Nc)_{k - \frac{\Nc}{2} + \frac{\Nf}{2} , k + \frac{\Nc}{2} + \frac{\Nf}{2}} 
       \Longleftrightarrow 
       U\left(  k - \frac{\Nc}{2} + \frac{\Nf}{2} \right)_{-\Nc,-k - \frac{\Nc}{2} - \frac{\Nf}{2}},
    \end{equation}
then from Table \ref{elec} and Table \ref{mag} we see that we are able to recover the IR TQFT around the SUSY point in the electric theory from the magnetic dual.

Notice that the IR dualities hold when in the electric theory we have $m_{\lambda}<0$ and $m_{\psi}>0$, while on the magnetic theory the mass have the opposite signs, i.e. $m_{\tlambda}>0$ and $m_{\tpsi}<0$ as expected. Since the SUSY point is located at $m_{\lambda}=-g^{2}k$ and $m_{\psi}=g^{2}k$ we claim that the IR dualities hold in a neighbourhood of the SUSY point. 

In what follows we will perform a similar analysis of the IR theories for non-SUSY theories and theories with different gauge groups. There we will not present all the possible TQFTs in the IR but rather limit ourselves to show the points in which the duality holds.

\section{Non-SUSY Dualities}

Now we consider the case $\tilde N_c <0$, where supersymmetry is spontaneously broken. When swapping the fivebranes we obtain $k-\frac{\Nc}{2}+\frac{\Nf}{2}$ \textit{anti}-$D3$-branes, therefore breaking supersymmetry\footnote{The appearance of anti branes in a 3D theory with a spontaneously broken supersymmetry was first pointed out in \cite{Maldacena:2001pb}.}. Then, the magnetic theory on the branes is
    \begin{equation}
        U\left(-k+\frac{\Nc}{2} -\frac{\Nf}{2}\right)_{\frac{k}{2} + \frac{3\Nc}{4} -\frac{\Nf}{4}, k+\frac{\Nc}{2} } + \Nf \, \tPhi.
    \end{equation}

Similarly, the magnetic' theory is $U\left(k + \frac{\Nc}{2} - \frac{\Nf}{2}\right)_{\frac{k}{2} - \frac{3\Nc}{4} +\frac{\Nf}{4}, k-\frac{\Nc}{2}}$ with $\Nf$ flavours. Note that $\tlambda$ is the Goldstino associated with the breaking of supersymmetry in the electric side.

We proceed to study the low energy dynamics of both sides of the duality. For $\Nf=0$ it was shown in \cite{Gomis:2017ixy} that the electric theory admits three phases: in addition to the large positive mass and large negative masses, there is an intermediate quantum phase characterised by a TQFT. This phases were recovered in \cite{Armoni:2022wef} using brane dynamics. Here we extend this dualities to the case including flavours and show that it also admits an intermediate phase.

As anticipated,  in this case the electric and electric' theories share the same range for $k$ (see footnote 1) . Since the theories cover the same regime, we conjecture that the $U(1)$ factor of both theories should decouple, thus, the electric and electric' theories are the same: $SU(\Nc)_{k}$. This theory has two magnetic duals. In order for them to correctly reproduce the IR dynamics of the electric theory, we need to ensure that their $U(1)$ levels match in the IR: we have to assign $U(1)$ level $\Nc-\Nf$ to the magnetic theory and $-\Nc+\Nf$ to the magnetic'. 

The matching of the low energy dynamics goes as follows: for the magnetic theory we integrate out $\tlambda$ and $\tpsi$ for $m_{\tlambda}>0$ and $m_{\tpsi}>0$, obtaining $U\left(-k+\frac{\Nc}{2}-\frac{\Nf}{2}\right)_{\Nc,\Nc}$, which is dual\footnote{Level-Rank duality for $SU$-$U$ groups is  given by \cite{Hsin:2016blu}
    \begin{equation*}
        SU(N)_{\pm K} \Leftrightarrow U(K)_{\mp N, \mp N}.
    \end{equation*}} 
to $SU(\Nc)_{k-\frac{\Nc}{2}+\frac{\Nf}{2}}$, while $m_{\tlambda}<0$ and $m_{\tpsi}<0$ leads to  $U\left(-k+\frac{\Nc}{2}-\frac{\Nf}{2}\right)_{k+\frac{\Nc}{2}-\frac{\Nf}{2}, \Nc-\Nf}$. These theories cover the negative mass and the intermediate phase in the electric theory. The magnetic' covers the intermediate phase and the positive mass regime. The details are as follows: for $m_{\tlambda}>0$ and $m_{\tpsi}>0$ we obtain $U\left(k + \frac{\Nc}{2} - \frac{\Nf}{2}\right)_{k-\frac{\Nc}{2}+\frac{\Nf}{2},-\Nc+\Nf}$ as TQFT in the IR, which is level-rank dual to $U\left(-k+\frac{\Nc}{2}-\frac{\Nf}{2}\right)_{k+\frac{\Nc}{2}-\frac{\Nf}{2}, \Nc-\Nf}$. For $m_{\tlambda}<0$ and $m_{\tpsi}<0$, we obtain $U\left(k + \frac{\Nc}{2} - \frac{\Nf}{2}\right)_{-\Nc,-\Nc}$ which is dual to $SU(\Nc)_{k+\frac{\Nc}{2}-\frac{\Nf}{2}}$. It is important to note that both $SU(\Nc)$ theories are the expected ones from the electric side when integrating out the adjoint fermion and the flavour (for opposite mass signs).

\section{\texorpdfstring{$USp$}{USp} and \texorpdfstring{$SO$}{SO} Theories}

In this section we generalise the previous analysis, to $USp(2\Nc)$ or $SO(2\Nc)$ (or $SO(2\Nc+1)$). We'll study both cases simultaneously. In order to realise these theories from the brane configuration we need to add an $O3^{+}$ ($O3^{-}$) plane extending in the 0126 directions. In the presence of the $O3^{+}$ we need to place $\Nc$ $D3$-branes and their mirrors since an odd number of branes is not compatible with the orientifold. On the other hand, when we place the $O3^{-}$ we can have an even or odd number of branes. This leads to the theories $USp(2\Nc)$ and\footnote{For the $SO$ case, $\Nc$ can be an integer of a half-integer} $SO(2\Nc)$. In both cases we will use a $(1,2k')$ tilted fivebranes and $2\Nf$ semi-infinite $D3$ flavour branes.

As in the $U(\Nc)$ case, when integrating out an adjoint fermion with mass $m$ the CS level is shifted as
    \begin{equation}
        k \rightarrow k + \frac{\text{sign}(m)}{2}h(G),
    \end{equation}
where $h(G)$ is the dual Coxeter number of the group $G$. In our case $G=USp(2\Nc)$ or $G=SO(2\Nc)$ we have $h(USp(2\Nc)) = 2\Nc+2$ and $h(SO(2\Nc)) = 2\Nc-2$.

\subsection{\texorpdfstring{$USp(2\Nc)$}{USp(2Nc)} Theory}

We start by setting $k' = k + \frac{1}{2}(\Nc+1) -\frac{\Nf}{2}$. On the electric side, we integrate out the adjoint fermion in the scalar multiplet for negative mass and one of the chiral flavours for positive mass. The resulting theory is $USp(2\Nc)_{2k}$ with $2\Nf$ flavours.  

On the magnetic side, when we swap the fivebranes an additional $D3$ anti-brane and its mirror image are created. After integrating out the adjoint fermion and one of the chiral flavours, we obtain the $\N=1$ theory $USp(2\tNc)_{-2\tilde{k}}$, with
    \begin{equation}
        \tNc = k' - \Nc + \Nf -1 = k - \frac{1}{2}(\Nc+1) + \frac{\Nf}{2}, \quad 2\tilde{k} = k + \frac{3\Nc}{2} - \frac{\Nf}{2} + \frac{1}{2}.
    \end{equation}

The magnetic dual is supersymmetric if $\tNc>0$. We propose the following $\N=1$ Seiberg duality
    \begin{equation}
        USp(2\Nc)_{2k} + 2\Nf \, \Phi \Leftrightarrow USp\left( 2\left(\frac{k}{2} - \frac{1}{2}(\Nc+1) + \frac{\Nf}{2}\right) \right)_{-2\left(\frac{k}{2}  + \frac{3\Nc}{4} - \frac{\Nf}{4} + \frac{1}{4}\right)} + 2 \Nf \, \tPhi.
    \end{equation}

The dynamics of the $USp$ theory goes as follows: for negative adjoint fermion mass and positive flavour mass the IR TQFT is $USp(2\Nc)_{2k -(\Nc+1)+\Nf}$. In the magnetic theory, for opposite sign masses with respect to the electric theory we obtain $USp( 2k-(\Nc+1)+\Nf)_{-2\Nc}$ TQFT. These theories agree due to level-rank duality\footnote{Level-rank duality for $USp$ CS theories is \cite{Aharony:2016jvv}
    \begin{equation*}
        USp(2N)_{2K} \Leftrightarrow USp(2K)_{-2N}
    \end{equation*}}
    \begin{equation}
        USp(2\Nc)_{2k -(\Nc+1)+\Nf} 
        \Leftrightarrow 
        USp( 2k-(\Nc+1)+\Nf)_{-2\Nc},
    \end{equation}

In addition, for small quark mass the magnetic squark is expected to condense resulting in a $USp(2\tNc-2n)$ SQCD theory and flavour symmetry breaking
    \begin{equation}
        USp(2\Nf) \rightarrow USp\left(2\Nf - 2n \right) \times USp\left(2n \right), 
    \end{equation}

described by the Grassmannian $USp(2\Nf)/USp\left(2\Nf - 2n \right) \times USp\left(2n \right)$, together with a WZ term, $2\Nc \int_{\mathcal{M}} \text{tr} \frac{1}{2\pi}F\wedge F$, with $F$ in the adjoint of $USp(2\Nf-2n)$. When we add SUSY breaking masses the selected vacuum is $n=\tNc$.

In the non-SUSY case, $\tNc<0$, again we have anti-$D3$-branes in the magnetic configuration. In this case the theory on the branes is still $USp(2\Nc)$ but the fermion transforms in the two-index antisymmetric representation, therefore when we integrate out the adjoint fermion the shift of the CS level is sign$(m_{\tlambda})(\tNc-1)$ instead of sign$(m_{\tlambda})(\tNc+1)$. With this consideration the magnetic dual is 
    \begin{equation}
        USp( -2k + (\Nc+1) - \Nf )_{2\tilde{k}} + 2\Nf \, \tPhi,
        \quad 2\tilde{k} = k + \frac{3\Nc}{2} - \frac{\Nf}{2} + \frac{1}{2},
    \end{equation}.
where we integrated out the fermions for $m_{\tlambda}>0$ and $m_{\tpsi'}>0$. Now we study the low energy dynamics of the magnetic theory. Integrating both the adjoint fermion and the flavour for positive mass we reach the TQFT 
    \begin{equation}
        USp( -2k + (\Nc+1) - \Nf )_{2\Nc},
    \end{equation}
    which corresponds to the large negative mass phase of the electric theory. When both fields have negative mass in the magnetic theory we obtain
    \begin{equation}
        USp( -2k + (\Nc+1) - \Nf )_{2k + \Nc -1+\Nf} 
        \Leftrightarrow
        USp( 2k+\Nc+1-\Nf )_{2k-\Nc+\Nf-1}
    \end{equation}
which covers the intermediate phase of the electric theory. Similar to the $U(\Nc)$ we can define the magnetic' theory 
    \begin{equation}
        USp( 2k+\Nc+1-\Nf)_{2\tilde{k}} + 2\Nf \, \tPhi, 
        \quad 2\tilde{k} = k - \frac{3\Nc}{2} + \frac{\Nf}{2} - \frac{1}{2},
    \end{equation}
which covers the intermediate and the large positive mass phases of the electric theory.

\subsection{\texorpdfstring{$SO(2\Nc)$}{SO(2Nc)} Theory}

The study of dualities for $SO(2\Nc)$ is almost identical to the $USp(2\Nc)$. We start by setting $k' = k + \frac{1}{2}(\Nc-1) -\frac{\Nf}{2}$. On the electric side we have $\N=1$ $SO(2\Nc)_{2k}$ with $2\Nf$ flavours. On the magnetic side, swapping the fivebranes a brane is created. After integrating out the adjoint fermion and one of the chiral flavours, we obtain the $\N=1$ theory $SO(2\tNc)_{-2\tilde{k}}$, with
    \begin{equation}
        \tNc = k' - \Nc + \Nf +1 = k - \frac{1}{2}(\Nc-1) + \frac{\Nf}{2}, \quad 2\tilde{k} = k + \frac{3\Nc}{2} - \frac{\Nf}{2} - \frac{1}{2}.
    \end{equation}

As before, the theory is SUSY if $\tNc>0$. We propose the following $\N=1$ Seiberg duality
    \begin{equation}
       SO(2\Nc)_{2k} + 2\Nf \, \Phi 
       \Leftrightarrow 
       SO\left( 2\left(-k - \frac{3\Nc}{2} + \frac{\Nf}{2} + \frac{1}{2}\right) \right)_{-2\left(\frac{k}{2} - \frac{3\Nc}{4} + \frac{\Nf}{4} + \frac{1}{4}\right)} + 2\Nf \, \tPhi 
    \end{equation}

In the IR, the TQFTs coincide when the adjoint fermion has negative mass and the flavour a positive one
    \begin{equation}
        SO(2\Nc)_{2k -(\Nc-1)+\Nf} 
        \Leftrightarrow
        SO( 2k-(\Nc-1)+\Nf)_{-2\Nc}
    \end{equation}
        
As before, we expect squark condensation for small quark. We obtain a $SO(2\Nc-2n)$ gauge theory flavour symmetry breaking pattern
    \begin{equation}
        SO(2\Nf) \rightarrow SO\left(2\Nf - 2n \right) \times SO\left(2n \right), 
    \end{equation}

described by the Grassmannian $SO(2\Nf)/SO\left(2\Nf - 2n \right) \times SO\left(2n \right)$, together with a WZ term $2\Nc \int_{\mathcal{M}} \text{tr} \frac{1}{2\pi}F\wedge F$, with $F$ in the adjoint of $SO(2\Nf-2n)$. When SUSY breaking masses are added the selected vacuum is $n=\tNc$.
    
In the non-SUSY case, $\tNc<0$, the magnetic theory is 
    \begin{equation}
        SO( -2k + (\Nc+1) - \Nf )_{2\tilde{k}} + 2\Nf \, \tPhi,
        \quad 2\tilde{k} = k + \frac{3\Nc}{2} - \frac{\Nf}{2} - \frac{1}{2},
    \end{equation}
where the adjoint fermion now transforms in the two-index symmetric representation because the theory lives on an anti-brane. Similarly the magnetic' theory is 
    \begin{equation}
        SO( 2k+\Nc-1-\Nf)_{2\tilde{k}} + 2\Nf \, \tPhi, 
        \quad 2\tilde{k} = k - \frac{3\Nc}{2} + \frac{\Nf}{2} + \frac{1}{2}.
    \end{equation}

In the IR, these theories reproduce the three phases of the electric theory.

\section{Conclusion}
 
In this paper we discussed the dynamics of non-Abelian ${\cal N}=1$ SQCD Chern-Simons theories. We used string theory to propose magnetic duals and using the magnetic description we were able to explore the IR dynamics of the electric theories. The dualities are valid not only at the supersymmetric point, but also in its vicinity, where supersymmetry is softly broken.
 
Our analysis revealed known as well as new phenomena, using string dynamics. In particular for $\tilde N_c \ge 0$ we found, apart from the semi-classical large quark mass phase, a quantum phase with symmetry breaking.
 
An interesting novelty is the identification of a WZ term in the magnetic description of the gauge theory. We argued that when a $D3$-brane is suspended between a $D5$-brane and a tilted fivebrane there is a Chern-Simons term localised on the fivebrane. While the presence of this term was argued in \cite{Komargodski:2017keh}, it was overlooked in previous analysis in both field theory and string theory \cite{Armoni:2017jkl,Benini:2018umh,Choi:2018ohn}.
 
Our dualities include the regime $\tilde N_c < 0$, where Seiberg duality was assumed to be invalid. Nevertheless, we showed that the IR theory of the electric and the magnetic theories is the same.
 
It will be interesting to exploit the techniques we used in the current analysis in other cases. In particular to obtain a better understanding of the WZ term in other field theories. In addition, our technique suggests that 4D Seiberg duality can be understood even when $N_f < N_c$, namely that the Affleck-Dine-Seiberg runaway superpotential can be obtained from a magnetic dual.

\section*{Acknowledgments}

We are thankful to Amit Giveon, Zohar Komargodski, David Kutasov, Vasilis Niarchos and Adar Sharon for valuable comments. The work of R.S. is supported by STFC grant ST/W507878/1.

\end{document}